\documentclass[aps,prl,showpacs,floats,twocolumn,floats,superscriptaddress,floatfix]{revtex4}
\usepackage{bm}
\usepackage{natbib}
\usepackage{amssymb}
\usepackage{amsfonts}
\usepackage{amsmath}
\usepackage{color}
\usepackage{times}
\usepackage{verbatim}
\usepackage{graphicx}
\usepackage{graphics,epsfig}
\usepackage{theorem}
\usepackage{makeidx}
\usepackage{amsmath}
\usepackage{epic}\usepackage{amscd}
\usepackage{bbm}
\usepackage{xy}
\usepackage{amssymb}
\usepackage{epsfig}

\newcommand{\ud}{\,\mathrm{d}}

\def\gsim{\;\rlap{\lower 2.5pt
\hbox{$\sim$}}\raise 1.5pt\hbox{$>$}\;}
\def\lsim{\;\rlap{\lower 2.5pt
\hbox{$\sim$}}\raise 1.5pt\hbox{$<$}\;}
\begin{document}

\newif\iffigs 
\figstrue
\iffigs \fi
\def\drawing #1 #2 #3 {
\begin{center}
\setlength{\unitlength}{1mm}
\begin{picture}(#1,#2)(0,0)
\put(0,0){\framebox(#1,#2){#3}}
\end{picture}
\end{center} }

\title{Upper Bound of 0.28eV on the Neutrino Masses from the Largest Photometric Redshift Survey}
\author{Shaun A. Thomas}
\author{Filipe B. Abdalla}
\author{Ofer Lahav}
\affiliation{Department of Physics and Astronomy, University College London, Gower Street, London, WC1E 6BT, UK}  

\begin{abstract}
We present a new upper limit of $\sum m_{\nu} \le 0.28$ ($95\%$ CL) on the sum of the neutrino masses assuming a flat $\mathrm{\Lambda CDM}$ cosmology. This relaxes slightly to $\sum m_{\nu} \le 0.34$ and $\sum m_{\nu} \le 0.47$ when quasi non-linear scales are removed and $w \ne -1$, respectively. These  bounds are derived from a new photometric redshift catalogue of over 700,000 Luminous Red Galaxies (MegaZ DR7) with a volume of 3.3 (Gpc $h^{-1}$)$^3$, extending over the redshift range $0.45 < z < 0.65$ and up to angular scales of $\ell_{\mathrm{max}} = 300$. The data are combined with WMAP 5-year CMB fluctuations,  Baryon Acoustic Oscillations (BAO), type 1a Supernovae (SNe) and an HST prior on the Hubble parameter. This is the first combined constraint from a photometric redshift catalogue with other cosmological probes. When combined with WMAP this data set proves to be as constraining as the addition of all SNe and BAO data available to date. The upper limit is one of the tightest and `cleanest' constraints on the neutrino mass from cosmology or particle physics. Furthermore, if the aforementioned bounds hold, they all predict that current-to-next generation neutrino experiments, such as KATRIN, are unlikely to obtain a detection.
 
\end{abstract}

\maketitle

{\it Introduction} -- Studies of the neutrino have traditionally been the realm of particle physics experiments, with Super-Kamiokande \citep{Fukuda98} first indicating the presence of mass. Neutrinos were shown to oscillate between the known flavors ($\nu_{e}$, $\nu_{\mu}$, $\nu_{\tau}$) solving, in the process, the long standing solar neutrino problem. This implies the neutrinos have at least two non-zero mass eigenstates ($m_{1}$, $m_{2}$, $m_{3}$) because the flavor mixing depends on the differences between their masses squared. Subsequently, bounds have been placed on the splitting {\it between} the neutrino mass eigenstates from a host of solar, accelerator and atmospheric experiments; $|\Delta m_{31}^{2}| \approx 2.4 \times 10^{-3} \mathrm{eV}^{2}$ and $\Delta m_{21}^{2} \approx 7.7 \times 10^{-5} \mathrm{eV}^{2}$ (e.g. \citep{Schwetz08}). However, currently both the absolute scale and the hierarchy of the masses remain hidden. KATRIN, a kinematic beta decay experiment \citep{Wolf08}, aims to provide a constraint in the future.

Cosmology not only probes the absolute mass scale of the neutrino but is a completely independent method to test against e.g. \citep{Elgaroy05,Lesgourgues06}. In any case, it is imperitive to include an accurate prescription for the neutrino in cosmology, as any failure to do so can bias the other cosmological parameters. A cosmological constraint on the sum of the neutrino masses is primarily a constraint on the relic Big-Bang neutrino density $\Omega_{\nu}$. One can relate this density to the sum of the mass eigenstates $\sum m_{\nu}$ (e.g. \citep{Dodelson03}) as given by,
\begin{equation} \label{eq:neutrinodensity}
\Omega_{\nu} = \frac{\sum m_{\nu}}{93.14 \mathrm{h}^{2} \mathrm{eV}}.
\end{equation}
\noindent
The direct effects of the neutrinos depend on whether they are relativistic, non-relativistic and the scale under consideration. Neutrinos have a large thermal velocity as a result of their low mass and subsequently erase their own perturbations on scales smaller than the \emph{free streaming} length \citep{Hu98b,Lesgourgues06}. This subsequently contributes to a suppression of the statistical clustering of galaxies over small scales and can be observed in a galaxy survey. The abundance of neutrinos in the Universe can also have a \emph{direct} effect on the primary CMB anisotropies if non-relativistic before the time of decoupling (i.e. when sufficiently massive). However, one of the most clear effects at this epoch is a displacement in the time of matter-radiation equality. All these cosmological effects can be used to impose bounds on the neutrino mass.

Previous studies have capitalised on these signatures and have started to place sub-eV constraints on the absolute mass scale \citep{Elgaroy02,Seljak06,Tereno08,Ichiki09,Komatsu09,Reid09}. We utilise the {\it new} SDSS MegaZ LRG DR7 galaxy clustering data \citep{Thomas09b} to provide the first photometric galaxy clustering constraint on the neutrino and, combining with the CMB, examine the complementarity of these early and late-time probes. With an almost comprehensive combination of probes this renders one of the tightest constraints on the neutrinos in cosmology and therefore physics.

{\it Assumptions} -- We assume a flat Universe with Gaussian and adiabatic primordial fluctuations and a constant spectral index. The effective number of neutrinos are fixed to $\mathrm{N_{eff}}=3.04$, thereby assuming no sterile neutrinos or other relativistic degrees of freedom. The constant dark energy equation of state is at first set to $w = -1$ and later relaxed. Finally, we consider the neutrinos to be completely mass degenerate given that current inferred bounds are much greater than the splitting hierarchies. The potential of future surveys to discriminate the mass hierarchy has been discussed in \citep{Lesgourgues04,Slosar06,Abdalla07,Kitching08,Bernardis09,Jimenez10}. 

{\it Analysis} -- Although parameter degeneracies and a mild insensitivity to relativistic (lighter) neutrinos limit the upper bound one can place on $\sum m_{\nu}$ with the CMB \citep{Ichikawa05} it represents a clean and relatively systematic-less cosmological tool whose high statistical discrimation of the remaining cosmological model facilitates a competitive combination of probes. We therefore start by using the latest 5-year WMAP data and likelihood as described in \citep{Dunkley08} to vary seven $\mathrm{\Lambda CDM}$ parameters: $\Omega_{b} h^{2}$, $\Omega_{c} h^{2}$, $\Omega_{\Lambda}$, $n_{s}$, $\tau$, $\mathrm{ln} (10^{10} A_{s})$ and $A_{SZ}$, in addition to $\sum m_{\nu}${\bf .} $\tau$, $n_{s}$ and $A_{s}$ represent the optical depth to reionisation, the scalar spectral index and the amplitude of curvature pertubations defined at $k=0.002 / \mathrm{Mpc}$, respectively. The contributions from the Sunyaev-Zeldovich fluctuations are included by adding a template spectrum $C_{\ell}^{SZ}$ with pre-factor $A_{SZ}$ following \citep{Komatsu02}. This is allowed to vary as $0 < A_{SZ} < 2$ \citep{Dunkley08}. We use the pre-March 2008 version of CAMB \citep{Lewis00} to produce the CMB power spectra. The reionisation is therefore treated as a semi-instantaneous process. The gravitational lensing effect on the CMB is also included, e.g. \citep{Seljak96}, and the {\sc CosmoMC} package \citep{Lewis02} is used for parameter exploration. 

\begin{figure*}
  \begin{flushleft}
    \centering
    \begin{minipage}[c]{1.00\textwidth}
      \centering
      \includegraphics[width=4.4cm,height=4.5cm]{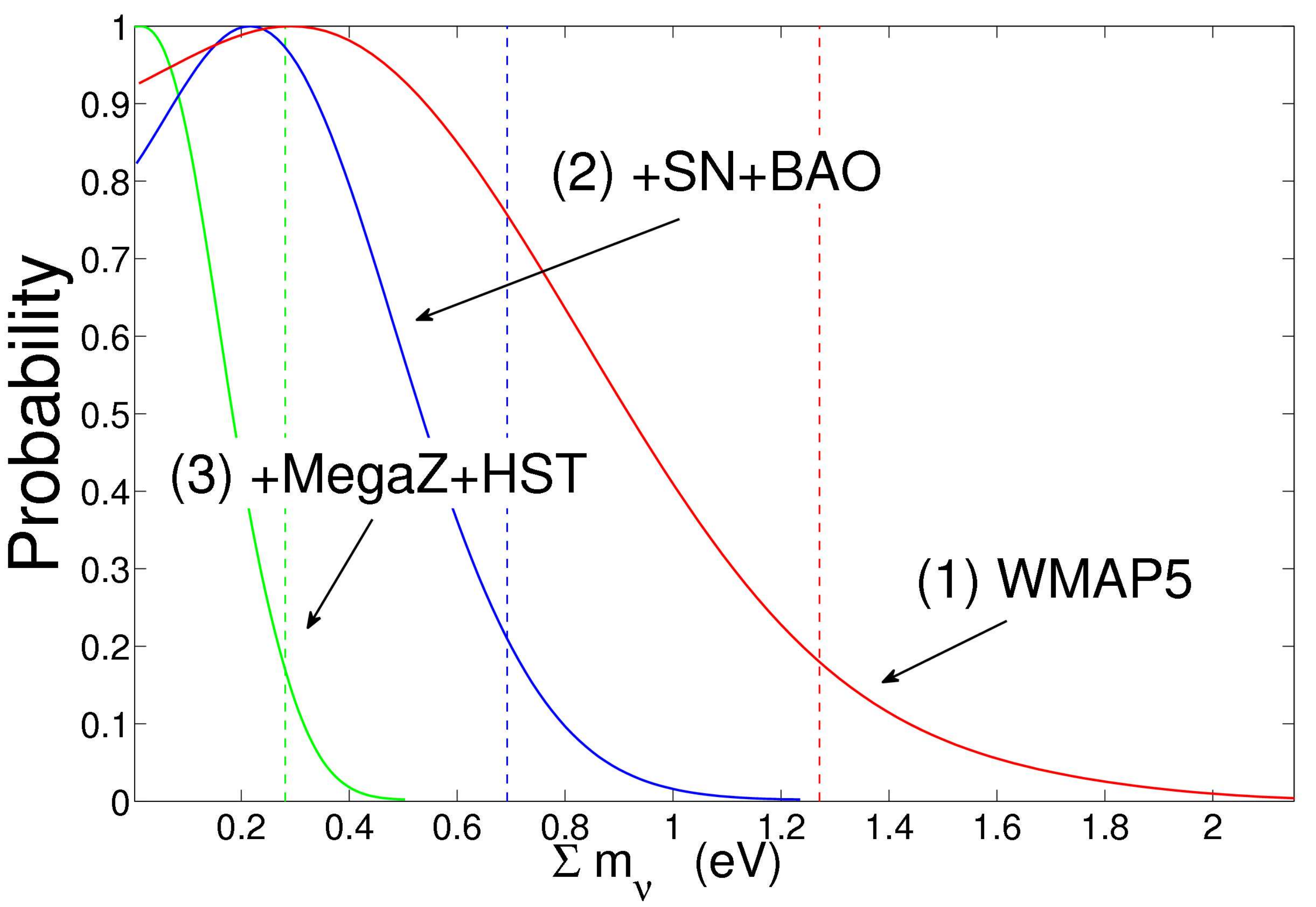}
      \includegraphics[width=4.43cm,height=4.5cm]{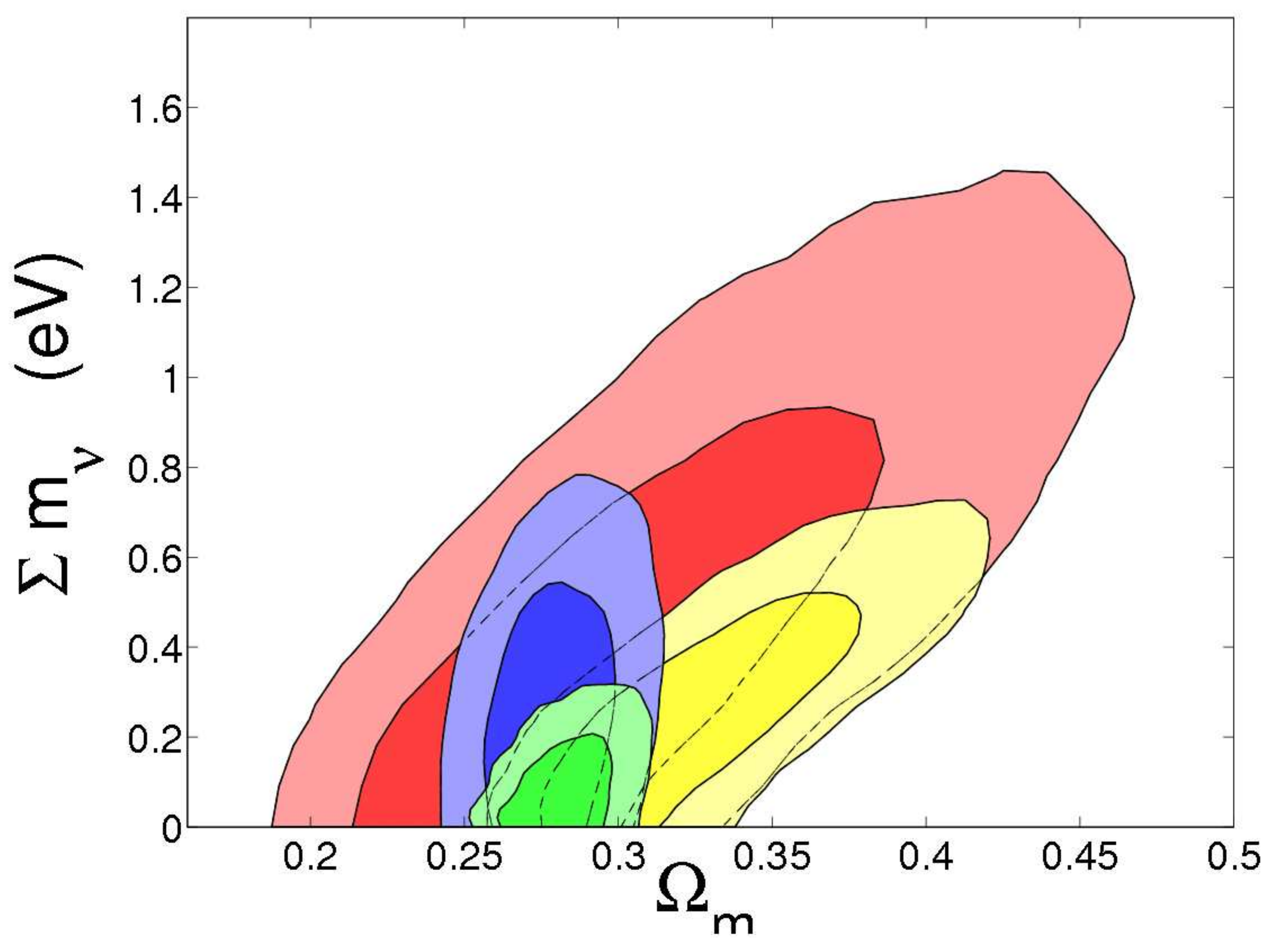}
      \includegraphics[width=4.43cm,height=4.5cm]{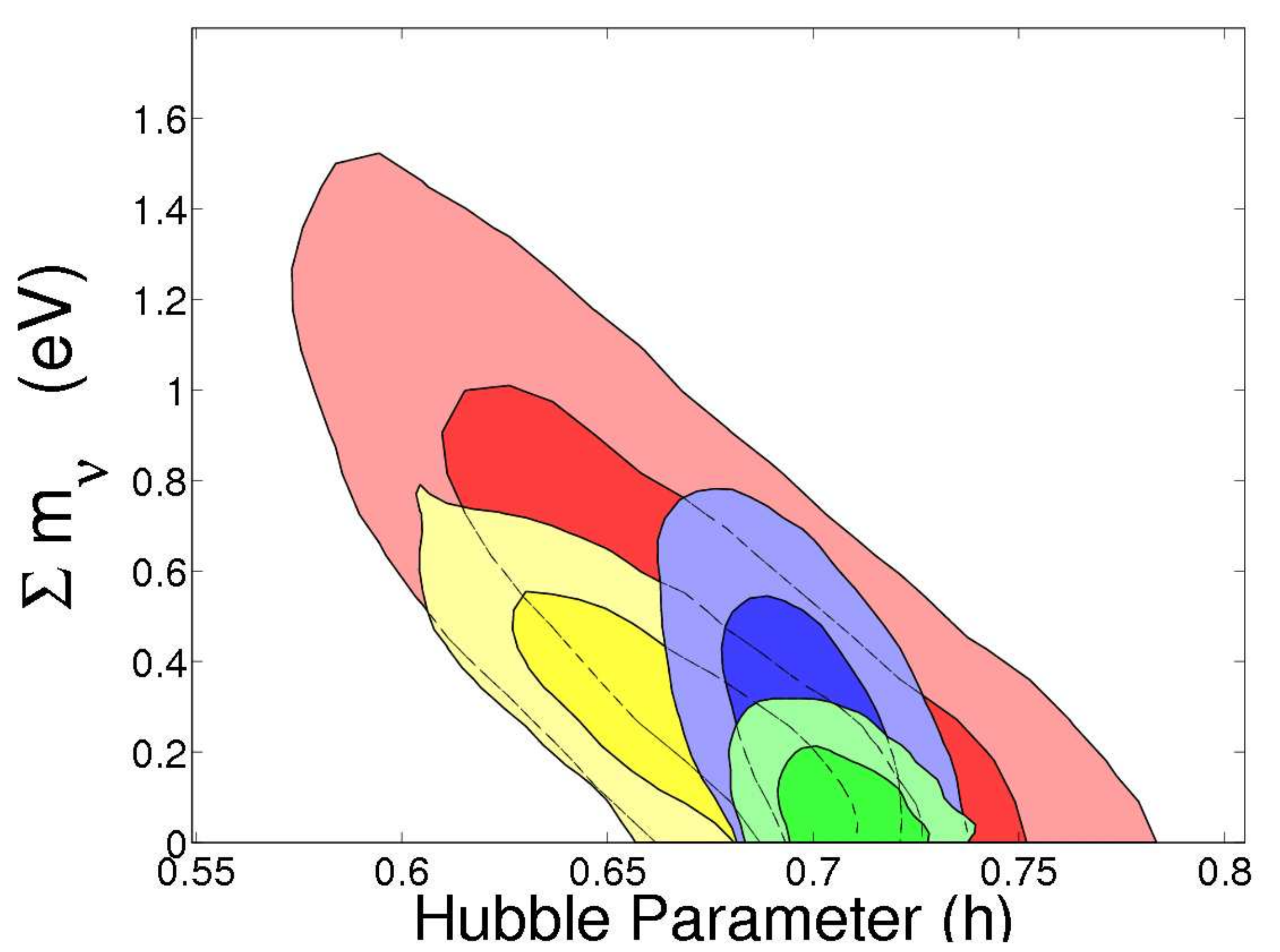}
      \includegraphics[width=4.43cm,height=4.5cm]{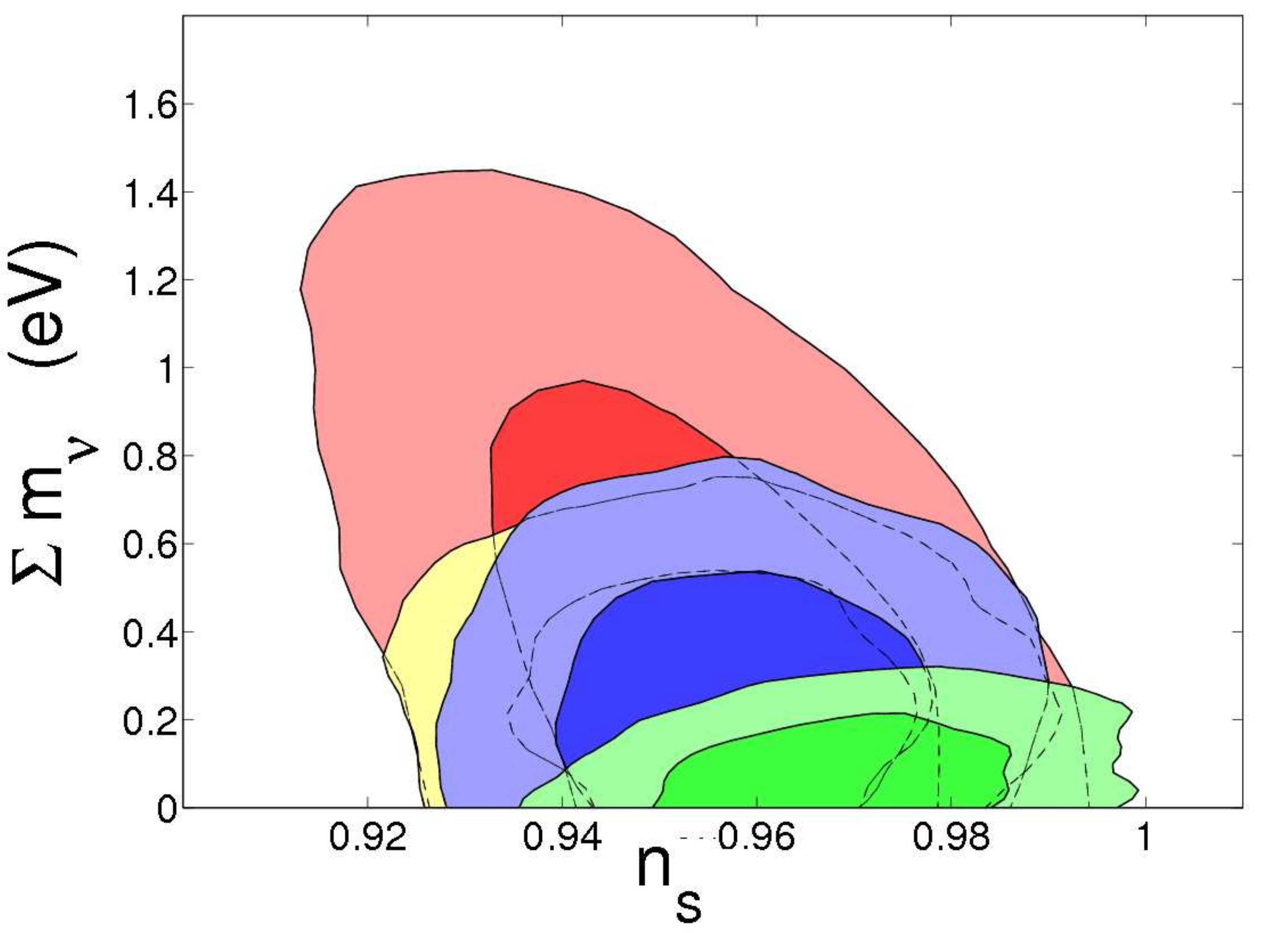}
    \end{minipage}
    \caption{\small{{\it Left Panel:} The marginalised 1D distribution for the neutrino from three incrementally combined analyses. The vertical dashed lines correspond to $95\%$ confidence levels. {\it Other Panels:} $68\%$ and $95\%$ marginalised distributions for the matter density $\Omega_{m}$, Hubble parameter $h$ and spectral index $n_{s}$ against the total neutrino masses. The contours correspond to (from bottom layer) WMAP-only (red/dark), WMAP+MegaZ (yellow/light), WMAP+SNe+BAO (blue/dark) and WMAP+SNe+BAO+MegaZ+HST (green/light). Table~\ref{table:neutrinosummary} highlights that the vast majority of the gain in the last two contours originates from the new and complementary galaxy clustering data.}}
    \label{fig:marginalisedcontours}
  \end{flushleft}
\end{figure*}

Our CMB run yields $\sum m_{\nu} < 1.271$ $\mathrm{eV}$ at the $95\%$ confidence level consistent with \citet{Komatsu09}. This bound implies the neutrinos were relativistic at decoupling and as such induces a degeneracy between the neutrino masses and $\Omega_{m}$ as well as the Hubble parameter $h$. This can be seen in Figure~\ref{fig:marginalisedcontours} as well as \citep{Ichikawa05,Komatsu09,Ichiki09}. This degeneracy can be improved by adding supernovae (SNe) data from the first year Supernova Legacy Survey (SNLS \citep{Astier06}) and the BAO data from \citep{Percival07}. Our analysis for WMAP + SNe + BAO gives $\sum m_{\nu} < 0.695$ eV ($95\%$ CL) similar to \citet{Komatsu09} ($\sum m_{\nu} < 0.67$ eV) and \citet{Ichiki09} ($\sum m_{\nu} < 0.76$ eV).

\begin{figure}
      \includegraphics[width=8.4cm,height=8.4cm]{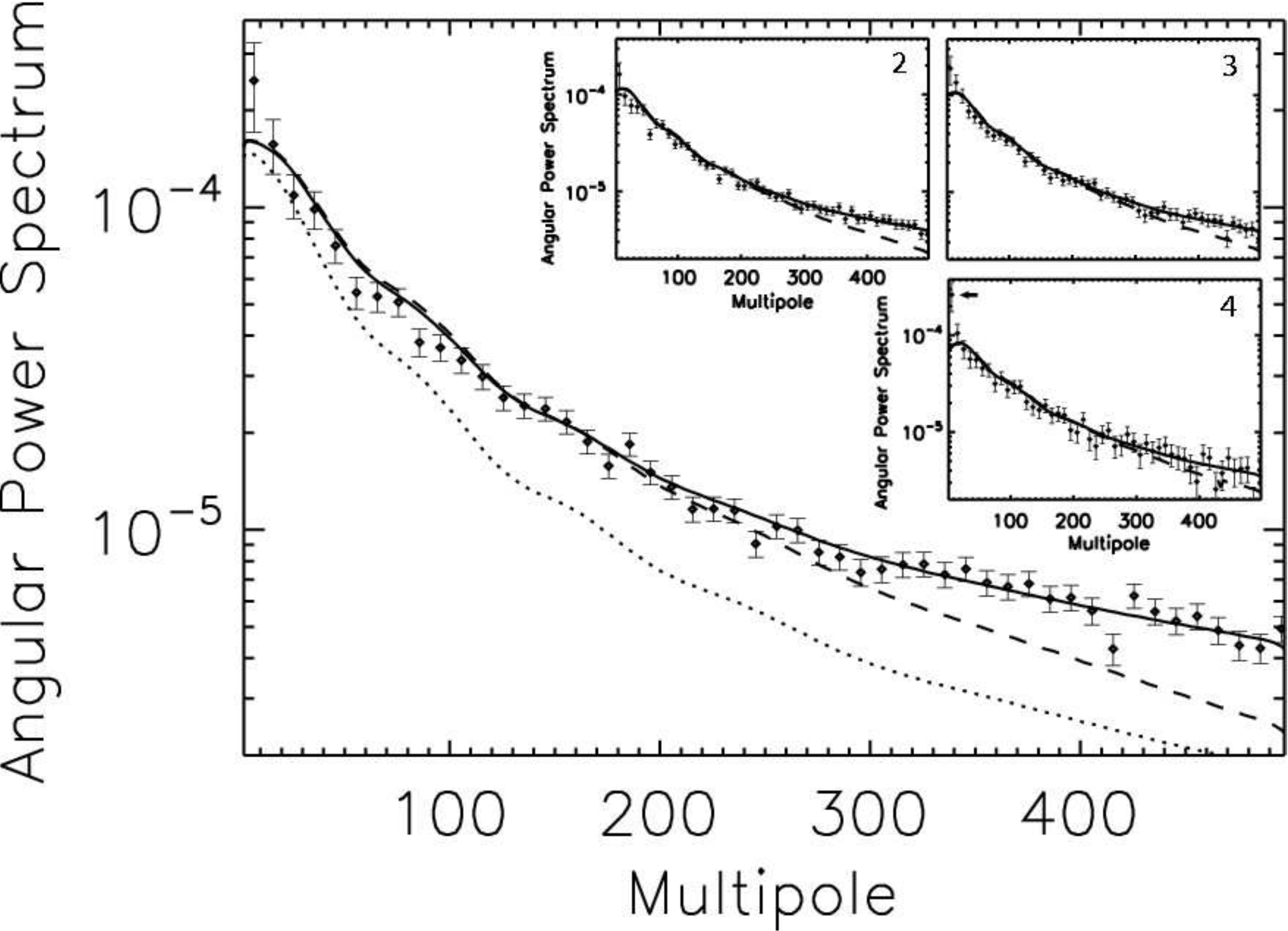}
    \caption{\small{The best fit angular power spectra $C_{\ell}$ in the combined analysis (solid lines) are plotted over the MegaZ DR7 data. The panels relate to four redshift bins with width $\Delta z = 0.05$ from $z = 0.45$ (main panel) to $z=0.65$ (panel 4). These spectra are good fits to the galaxy statistics including scales not utilised in the analysis ($\ell > 300$). The fit is also plotted for linear spectra (dashed lines) and highlights the scale at which the non-linear regime starts to become significant. The dotted line demonstrates the effect of introducing $\sum m_{\nu} = 1$ eV neutrinos with all parameters, except $\Omega_{c}$, held fixed.}}
    \label{fig:neutrinobestfitdata}
\end{figure}
\noindent
In order to go beyond such studies we include the MegaZ LRG (DR7) \emph{photometric} redshift survey that will be presented in \citep{Thomas09b}, which we have checked to be compatible with earlier SDSS clustering \citep{Blake07} and photo-z analyses \citep{Abdalla08,Collister04}. This adds galaxy clustering information that is sensitive to the growth of structure suppressed by the free streaming neutrinos. The SDSS colours provide reliable photometric redshift estimates and, due to their high luminosity, probe a large region of cosmic volume. Encapsulating $7746$ $\mathrm{deg}^{2}$,  we utilise 723,556 photometrically determined LRGs in four redshift bins of width $\Delta z = 0.05$ between $0.45<z<0.65$ in a spherical harmonic analysis of the galaxy distribution until a maximum multipole $\ell_{max}=300$. These galaxies are calibrated by the 2SLAQ redshift survey \citep{Cannon06} using ANNz \citep{Collister04} as in \citep{Abdalla08} and the previous DR4 photometric release \citep{Blake07,Collister07}. Specifically we use the angular power spectrum defined as, 

\noindent
\begin{equation} \label{eq:angularpowerspectrum}
C_{\ell} \equiv <\delta^{2D} \delta^{* 2D}> = 4\pi \int \Delta^{2}(k)W_{\ell}^{2}(k) \frac{\ud k}{k}.
\end{equation}
\noindent where $\Delta^{2}(k)$ is the dimensionless power spectrum calculated with {\sc CAMB}. The matter distribution is projected onto a plane in the sky with weight $W_{\ell}^{2}(k)$ in this statistic described by both $W_{l}(k) = \int f(z)j_{l}(kz) \ud z$ and $f(z) = n(z) D(z) (\frac{\ud z}{\ud x})$, with the spherical Bessel function $j_{l}(kz)$, the linear growth factor $D(z)$ and the normalised redshift distribution $n(z)$. The effects of redshift space distortions are included as described in \citep{Fisher94,Padmanabhan07}. The likelihood combines the four measured redshift bins and includes the full covariance as a result of photometric errors scattering galaxies between bins and therefore correlating slices. There are four additional parameters included in the study as a result of the galaxy bias in each bin ($b_{1}$, $b_{2}$, $b_{3}$ and $b_{4}$), i.e. modestly accounting for the redshift dependence, which is seen to increase at high $z$ \citep{Blake07,Thomas09b}. Despite the non-linear contribution becoming significant only at scales $\ell > 300$ we use {\sc halofit} \citep{Smith03} to model the non-linear power spectrum. It is interesting to note that the point corresponding to the largest angular scale band in the highest redshift bin indicates an excess of power. Hints of this were seen in the earlier photometric releases by \citep{Blake07,Padmanabhan07} with the excess labelled by the arrow in panel 4 in Figure~\ref{fig:neutrinobestfitdata}.

This survey is not only one of the most recent and largest to date but is one of the most competitive available. However, these power spectra provide an additional incentive for this combined measurement. This because the BAOs, which were shown to be so advantageous before, can be used in conjunction to MegaZ with no cross-covariance. The BAO data is extracted at $z=0.2$ and $z=0.35$, whereas MegaZ is defined at a higher redshift. They therefore constitute two independent data sets and can be used both simply and \emph{simultaneously}. 

By combining the MegaZ LRGs as described above with the previous CMB, SNe and BAO data in a complete joint analysis we find a significantly lower bound of $\sum m_{\nu} < 0.325$ eV at the $95\%$ confidence level. Again, this is roughly a factor $2$ improvement in the neutrino masses with the addition of the LRGs and is shown clearly against $\Omega_{m}$, $h$ and the 1D marginalised distribution in Figure~\ref{fig:marginalisedcontours}.

The information on the growth of stucture is paramount to the improvement seen in this study. However, part of this information originates from the quasi-non-linear regime and could systematically bias the inferred constraint. While approaches are developed and work continues into the effects of the neutrino on these scales (E.g. \citep{Saito09}) we repeat the combined analysis with the smaller scales removed. By truncating the multipoles at $\ell_{\mathrm{max}} = 200$ this more conservative approach is seen to give a similar but slightly relaxed limit of $\sum m_{\nu} < 0.393$ eV. While this highlights the importance of understanding non-linearities for obtaining the most stringent constraints, it is reassuring that there is still a marked improvement on the previous study (CMB+SNe+BAO) with linear LRGs.

It is also intriguing to compare the input of the LRGs to those of the two distance measures (SNe+BAO). These have previously been highly beneficial to the uncertainty. We therefore perform a joint analysis using just the WMAP5 and LRG data, subsequently obtaining the limit $\sum m_{\nu} < 0.651$ eV at the $95\%$ confidence level. This is comparable to the \emph{spectroscopic} DR7 galaxy clustering addition to the CMB in \citep{Reid09} with $\sum m_{\nu} < 0.62$ eV and illustrates the development of photometric surveys as a competitive tool for the future.

\begin{figure}
      \includegraphics[width=8.2cm,height=8.0cm]{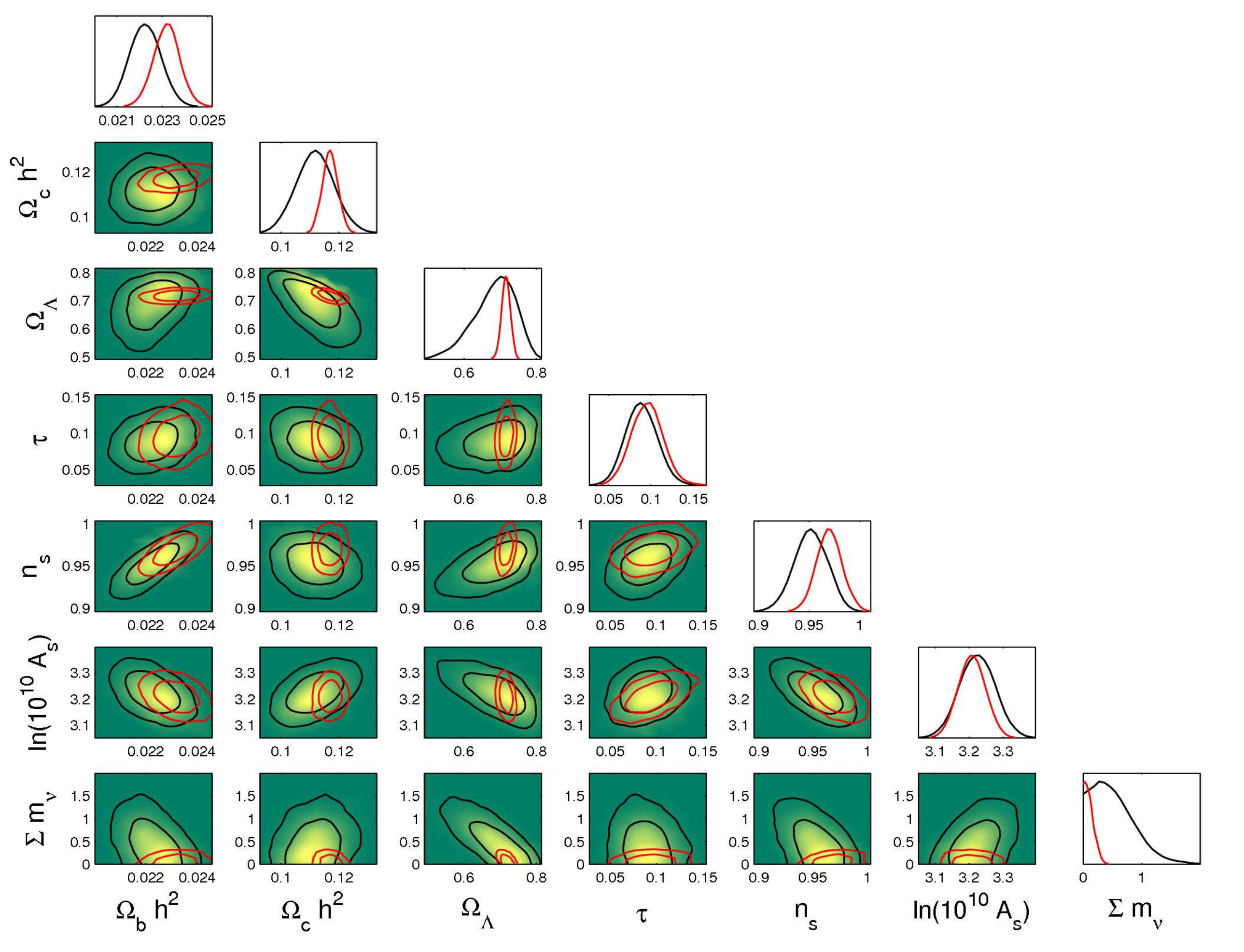}
    \caption{\small{The 2D $68\%$ and $95\%$ contours and marginalised 1D distributions for 7 cosmological parameters ($\Omega_{b} h^{2}$, $\Omega_{c} h^{2}$, $\Omega_{\Lambda}$, $n_{s}$, $\tau$, $\mathrm{ln} (10^{10} A_{s})$ and $\sum m_{\nu}$) in a WMAP5 + SNe + BAO + MegaZ DR7 + HST combined constraint (red contours). The amplitude of the Sunyaev-Zeldovich fluctuations ($A_{SZ}$) and four bias parameters ($b_{1}$, $b_{2}$, $b_{3}$, $b_{4}$) are marginalised over but not plotted. The black contours are given for the WMAP-only analysis. }}
    \label{fig:cmb_sn_megaz_neutrino_dr7}
\end{figure}

We conclude this work by further restricting the cosmological parameter space with the addition of the new HST prior on the Hubble parameter to the WMAP5 + SNe + BAO + MegaZ DR7 run. The improved prior was recently found to be: $H_{0} = 74.2 \pm 3.6$ $\mathrm{km \; s^{-1} Mpc^{-1}}$ \citep{Riess09b}. With this the final limit in this study is reduced to $\sum m_{\nu} < 0.28$ eV at the $95\%$ confidence level. This is one of the tightest constraints in the literature. The angular power spectra $C_{\ell}$ corresponding to the best fit values are plotted in Figure~\ref{fig:neutrinobestfitdata} with the galaxy clustering data. An overview of all the neutrino bounds are displayed in Table~\ref{table:neutrinosummary} and a plot of all parameter combinations compared to the CMB-only study is displayed in Figure~\ref{fig:cmb_sn_megaz_neutrino_dr7}. Our estimates on the bias are $b_{1} = 1.74 \pm 0.07$, $b_{2} = 2.02 \pm 0.08$, $b_{3} = 2.12$ $\pm$ $0.09$ and $b_{4} = 2.39$ $\pm$ $0.10$. 

For $w \ne -1$ the tighter bound relaxes slightly to $\sum m_{\nu} < 0.47$ eV, which should be compared to $\sum m_{\nu} \lesssim 0.62$ eV from \citep{Goobar06}. We note that biasing \emph{could} act to mimic the neutrino signature over smaller scale analyses. As a gauge of this effect we implement, as an example, the `Q-model' of \citep{Cole05}; resulting in a combined constraint (all data) of $\sum m_{\nu}$ $< 0.44$ eV. However, undertaking the challenge of modelling the possibility of scale dependent bias accurately and self consistently is left to future work.

\begin{table}
\center
\begin{tabular}{ll} 
\\
\hline
$\sum m_{\nu}$ ($95\%$ CL) & Analysis \\
\hline
\hline
$ < 1.271$ eV & WMAP5 \\
$ < 0.695$ eV & WMAP5 + SNe + BAO \\
$ < 0.651$ eV & WMAP5 + MegaZ\\
$ < 0.393$ eV & WMAP5 + SNe + BAO + MegaZ$_{(\ell_{\mathrm{200}})}$\\
$ < 0.344$ eV & WMAP5 + SNe + BAO + MegaZ$_{(\ell_{\mathrm{200}})}$ + HST\\
$ < 0.325$ eV & WMAP5 + SNe + BAO + MegaZ \\
$ < 0.281$ eV & WMAP5 + SNe + BAO + MegaZ + HST\\
\hline
$ < 0.491$ eV & WMAP5 + SNe + BAO + MegaZ$_{(\ell_{\mathrm{200}})}$ + HST \\
$ < 0.471$ eV & WMAP5 + SNe + BAO + MegaZ + HST\\
\hline
\hline
\end{tabular}

\caption{\small{A summary of the bounds placed on $\sum m_{\nu}$ in this \emph{letter}. $\ell_{\mathrm{200}}$ corresponds to the truncation in the maximum multipole scale to remove the quasi non-linear regime. The top constraints are for $w=-1$; the bottom for $w \ne -1$, marginalised over.}}
\label{table:neutrinosummary}
\end{table}
{\it Conclusions} -- Using the biggest ever large scale structure survey we have set bounds on the neutrino masses at $\sum m_{\nu} < 0.28$ eV ($\ell_{\mathrm{max}}=300$) and $\sum m_{\nu} < 0.34$ eV ($\ell_{\mathrm{max}}=200$) at $95\%$ CL, when combined with WMAP5+SNe+BAO+HST data. This is the first ever determination of neutrino masses from a photometric galaxy redshift survey. Not only have we shown that photometric redshifts can be used for this problem, but also that such a galaxy survey is competitive with all currently available geometric probes (SNe+BAO) or spectroscopic clustering when added to the CMB. Our constraint is one of the tightest current bounds available without the use of data from Lyman-$\alpha$ (e.g. \citep{Seljak06}), which is prone to systematics, or a complicated modelling of the bias \citep{deBernardis08}. Further, all our results show that KATRIN's \citep{Wolf08} projected $90\%$ sensitivity ($\sum m_{\nu}$ $\lesssim 0.6$ eV) leaves an unlikely neutrino mass detection. Finally, our overall method shows great promise for the next generation of surveys, which will yield upper limits of e.g. $0.12$ eV \citep{Lahav09} and $0.025$ eV \citep{Abdalla07} at $95\%$ C.L.

{\it Acknowledgements} -- ST acknowledges a STFC studentship and UCL's Institute of Origins for a Post-doctoral Fellowship. FBA and OL acknowledge the support of the Royal Society via a Royal Society URF and a Royal Society Wolfson Research Merit Award, respectively. OL and FBA acknowledge the Weizmann Institute of Science for an Erna and Jakob visiting professorship and support for a research visit, respectively. 

\bibliography{paper}

\end{document}